\documentclass{article}
\usepackage{spconf,amsmath,graphicx}
\graphicspath{{figs/}}
\usepackage{color}

\usepackage{bm}

\def\c{{\bm c}}

\def\x{{\bm x}}
\def\z{{\bm z}}

\title{Deep Learning for Tube Amplifier Emulation}
\name{Eero-Pekka Damsk{\"a}gg, Lauri Juvela, Etienne Thuillier, and Vesa V{\"a}lim{\"a}ki}
\address{Acoustics Lab, Dept. of Signal Processing and Acoustics,\\
	Aalto University, Espoo, Finland\\
    email: firstname.lastname@aalto.fi}
\begin{document}
\onecolumn
{\noindent\Large \textbf{IEEE Copyright Notice}}

${}$

{\noindent \large \copyright 2019 IEEE.
Personal use of this material is permitted.  Permission from IEEE must be obtained for all other uses, in any current or future media, including reprinting/republishing this material for advertising or promotional purposes, creating new collective works, for resale or redistribution to servers or lists, or reuse of any copyrighted component of this work in other works.}

${}$
\twocolumn

\maketitle
\begin{abstract}
Analog audio effects and synthesizers often owe their distinct sound to circuit nonlinearities. Faithfully modeling such significant aspect of the original sound in virtual analog software can prove challenging. The current work proposes a generic data-driven approach to virtual analog modeling and applies it to the Fender Bassman 56F-A vacuum-tube amplifier. Specifically, a feedforward variant of the WaveNet deep neural network is trained to carry out a regression on audio waveform samples from input to output of a SPICE model of the tube amplifier. The output signals are pre-emphasized to assist the model at learning the high-frequency content. The results of a listening test suggest that the proposed model accurately emulates the reference device. In particular, the model responds to user control changes, and faithfully restitutes the range of sonic characteristics found across the configurations of the original device.
\end{abstract}

\section{Introduction}
\label{sec:intro}

Digital modeling of analog audio circuits has become an active research topic in recent years \cite{Karjalainen06, Yeh2009, Valimaki11}. Such virtual analog models are essential, when music technology is computerized and there is a need to produce all musical sounds by means of software. The distinct sonic characteristics of many musical devices, such as those of vacuum-tube amplifiers and distortion effects used by guitar and bass players, arise from their nonlinear behavior \cite{Barbour98, pakarinen2009review}. Digital models of such systems should emulate these nonlinear characteristics accurately in order to produce a pleasing and realistic sound.

There are two main approaches to virtual analog modeling: ``black-box'' and ``white-box'' modeling. White-box modeling is based on circuit analysis and simulation, such as wave digital filters (WDF) \cite{Karjalainen06, dunkel2016fender}. Such models exhibit a high degree of accuracy, but require detailed knowledge of the circuit and its nonlinear components. Additionally, these models often implement nonlinear solvers, which increases the computational load.

Black-box modeling is based on input-output measurements of the device under study. In virtual analog modeling, black-box modeling approaches include dynamical convolution \cite{kemp1999analysis, farina2005emulation}, the Volterra series \cite{helie2010volterra, orcioni2018identification}, as well as its special cases, the Hammerstein and Wiener models \cite{tronchin2013emulation}. Black-box modeling is appealing for its generality. In principle, a black-box model can be used to emulate different nonlinear systems, but it typically fails to emulate the system as accurately as a white-box model, leading to an audible difference between the model and the device under study. Furthermore, black-box models of nonlinear systems typically do not include user controls, and instead a separate model has to be estimated for each different configuration of the device under study.

In this work, these limitations of conventional black-box modeling methods are addressed using a deep neural network.
A feedforward variant of the WaveNet architecture \cite{wavenet2016oord} is trained for modeling of the vacuum-tube guitar preamplifier.
Neural networks have been applied recently for tube amplifier modeling by several authors \cite{covert2013vacuum, zhang2018vacuum, schmitz2018nonlinear}.
However, the previous models have not incorporated user controls, and are limited to a single setting of the amplifier. The WaveNet model proposed in this study can be conditioned with user control settings, allowing a single model to represent all settings. Furthermore, this paper presents results of a listening test showing that the proposed WaveNet model provides a better subjective quality than other methods tested.

The rest of this paper is structured as follows. Sec.~\ref{sec:bassman} discusses the modeled preamplifier circuit. Sec.~\ref{sec:model} details the proposed neural network architecture for black-box modeling of nonlinear audio circuits. In Sec.~\ref{sec:evaluation}, the proposed approach is evaluated and compared against other methods using an objective metric and a listening test. Sec.~\ref{sec:conclusions} concludes the paper.

\section{Device Under Study}
\label{sec:bassman}
The device modeled in this study is the preamplifier circuit from the Fender Bassman 56F-A vacuum-tube amplifier. The Fender Bassman was originally developed as a bass amplifier but found, including its many derivatives---such as the Marshall JTM~45---significant use among guitarists as well. The Bassman family of preamplifiers has been studied earlier in~\cite{dunkel2016fender} where a WDF-based white-box model is developed. The circuit contains four vacuum tube triodes in a complex topology that is challenging for white-box modeling. A more thorough circuit analysis is given in~\cite{dunkel2016fender}.

An important factor when modeling with a feedforward neural network lies in the duration of the original device's response to different stimuli. This gives an estimate for the number of past input samples to be shown the model for making an accurate prediction about the output, i.e. the feedforward model order. For a linear circuit, this is determined directly from the impulse response. However, in the case of a nonlinear circuit, the estimation of the required order is not straightforward as it depends on the input signal.

\begin{figure}[t]
\begin{minipage}[b]{1.0\linewidth}
  \centering
  \centerline{\includegraphics[width=8.5cm]{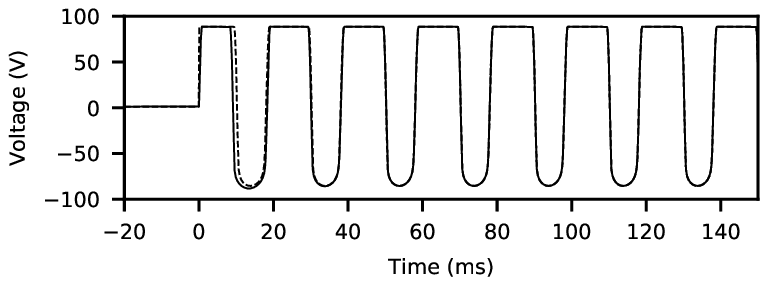}}
\end{minipage}
\begin{minipage}[b]{1.0\linewidth}
  \centering
  \centerline{\includegraphics[width=8.5cm]{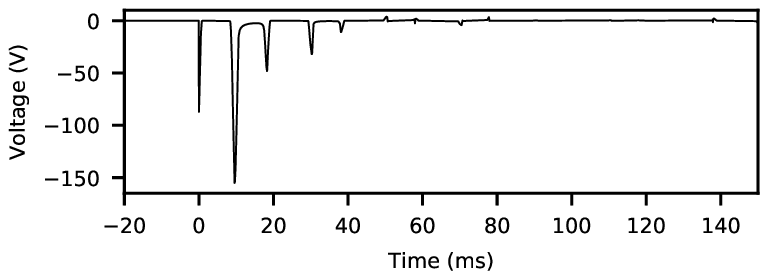}}
\end{minipage}
\caption{(Top) The circuit's response to a 50-Hz pure tone (solid) and the estimated steady-state response (dashed). The length of the transient response  (bottom) is used to determine the required input length for the neural network.}
\label{fig:response}
\end{figure}

In practice, we used several pure tone probe signals with various frequencies and amplitudes and extracted the corresponding transient responses by subtracting the steady-state response observed at $t \gg 1$\,s from the measured output. The output signal, estimated steady-state and transient responses are shown in Fig. \ref{fig:response} for a 50-Hz pure tone probe signal with a peak-to-peak amplitude of 0.5\,V. As suggested in this example, the transient response of the system becomes insignificant after about 80\,ms, such that the neural network should be able to accurately predict the next output sample, given a sliding window of the past input samples this long.
\section{Proposed model}
\label{sec:model}

The proposed model is a variant of the WaveNet convolutional neural network \cite{wavenet2016oord}. The WaveNet architecture has two main parts: a stack of dilated causal convolution layers and a fully-connected post-processing module. The convolutional layers form a nonlinear filter bank. The output of each convolutional layer is fed as input to the post-processing module, which is a fully-connected neural network. In the original WaveNet model, the post-processing module had a softmax output layer for representing the probability distribution of the next sample value in the sequence.

\begin{figure}[t]
\centering
  \centerline{\includegraphics[width=8.3cm]{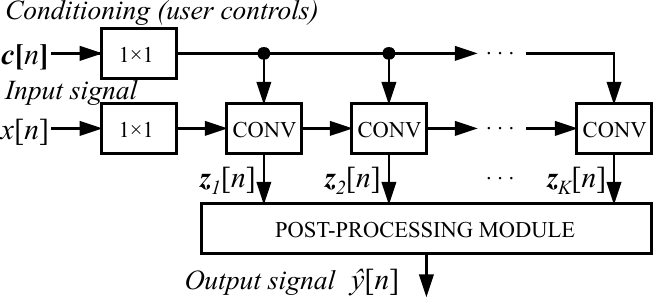}}
\caption{Proposed deep neural network architecture.}
\label{fig:model}
\end{figure}

In this work, we adapt the original WaveNet model for a regression task. The architecture is shown in Fig.~\ref{fig:model}. A similar architecture has been proposed for speech denoising~\cite{rethage2018wavenet}. Instead of modeling a conditional probability over the next sequence value, the model is trained to predict the current output sample value, given a certain number of past input samples and the current input sample:
\begin{equation}
    \hat{y}[n, \bm{\theta}] = f(x[n],\dots,x[n-N], \bm{\theta}),
\end{equation}
\noindent where $N$ is called the receptive field of the model, $f(\cdot)$ is a nonlinear transformation learned by the model, and $\bm{\theta}$ are the learned network parameters. In this work, the output $\hat{y}[n]$ corresponds to the audio signal processed through the model of the audio circuit, and $x[n]$ is the unprocessed input signal.

Conditioning is included in the model, allowing it to adapt according to user control adjustments. In case of the Fender Bassman preamp, the user can adjust the gain in the voltage amp stage by adjusting a potentiometer. To incorporate user controls to the model, the local conditioning procedure presented in the original WaveNet paper is used~\cite{wavenet2016oord}. The conditioning is realized in the convolutional layers, which use the following gated activation:
\begin{equation}
    \z = \textrm{tanh}(W_f * \x + V_f * \c)\odot \sigma(W_g *\x + V_g * \c),
\end{equation}
where $*$ and $\odot$ denote the convolution and the element-wise multiplication operations, respectively, $W_f$ and $W_g$ are the filter and gate convolution kernels, respectively, $\x$ is the input, and $\sigma(\cdot)$ is the logistic sigmoid function. The convolutions with the conditioning kernels, $V_f * \c$ and $V_g * \c$, where $\c$ is a sequence representing the user control values, are $1 \times 1$ convolutions, which means that each filter in $V_f$ and $V_g$ has length 1. The convolutional layers include residual connections so the input to the next convolutional layer is $\x_{k+1} = \x_k + W_k * \z_k$, where $W_k * \z_k$ is a $1 \times 1$ convolution, and $k$ is the layer index.

The inputs to the fully-connected post-processing module are the outputs $\z_k$ of the convolutional layers. The first layer of the post-processing module uses the gated activation function. The second layer has a hyperbolic tangent activation. Finally, a linear layer outputs the predicted sample value.

\subsection{Loss function}
The model parameters are learned by minimizing the ``error-to-signal ratio'' (ESR), which is the squared error divided by the energy of the target signal:
\begin{equation}
    \mathcal{E} = \frac{\sum_{n=-\infty}^{\infty} | y[n] - \hat{y}[n,\bm{\theta}] |^2}{\sum_{n=-\infty}^{\infty} | y[n] |^2}.
    \label{eq:loss}
\end{equation}
Additionally, a pre-emphasis filter is applied to the target and predicted signal before computing the error in (\ref{eq:loss}). This work has chosen to use the high-pass pre-emphasis filter ${H(z) = 1 - 0.95 z^{-1}}$, which gives more weight to the error at high frequencies. According to our tests, without the pre-emphasis filtering, the model struggles at high frequencies.

\subsection{Training}
The training data was generated by processing audio signals with a SPICE circuit model of the Fender Bassman preamplifier. In virtual analog literature, a SPICE model is commonly used as the ``ground truth'', as it is a convenient way of providing training signals, which are very similar to the data measured from a physical device. In this work, the SPICE model was implemented according to the circuit diagram presented in \cite{dunkel2016fender}, using the 12AX7 triode model proposed in~\cite{dempwolf2011physically}.

The clean signals used as input to the SPICE model were obtained from an audio database  released for a general-purpose audio tagging challenge \cite{fonseca2018general}. We randomly chose subsets of 4.0\,h for training and 20\,min for validation.
To generate the training and validation data, the unprocessed audio signals in the database were further split into 100-ms segments. A uniformly sampled random gain value between $-15$ and $15$\,dB was applied to each input segment to increase the dynamic range of the signals in the data. This was important, as the output signal sounds very different for different input levels. 
Each segment was then fed to the SPICE model to obtain the target data set. For each segment, the gain setting in the preamplifier was also sampled uniformly and stored alongside the input/target pair. 
The models were trained using the Adam optimizer \cite{diederik2014adam} with early stopping based on the validation error computed after each epoch.

\section{Evaluation}
\label{sec:evaluation}

For evaluation of the proposed model, it is compared with two other black-box models: a multilayer perceptron (MLP) and the block-oriented model proposed by Eichas and Z\"olzer~\cite{eichas2016black}. The MLP is included as a neural network baseline, and the block-oriented model is included as it represents the current state-of-the-art in black-box modeling of nonlinear audio systems. The inputs to the MLP are the current input sample and past $N$ samples, where $N$ is set equal with the receptive field of the WaveNets. In the hidden layers, the MLP uses hyperbolic tangent activations and the same additive conditioning procedure as the WaveNet: 
\begin{equation}
    \z = \textrm{tanh}(W\x + V\c).
\end{equation}

The block-oriented model \cite{eichas2016black} consists of a linear filter followed by a parametric nonlinearity. It was used for modeling the Bassman preamp in \cite{eichas2017block}. Here, the linear filters and the parameters of the nonlinearity are optimized jointly with the Adam optimizer, using a generated dataset consisting of sinusoids processed with the SPICE model. As the model does not support user control adjustments, separate models were trained for the two preamplifier gain settings tested. 

\subsection{Model configuration}
\label{sec:model-config}
The WaveNet model has several hyperparameters which affect the performance of the model.
In this work, it was decided to use ten convolutional layers with a filter width of three and the dilation pattern $d_k = \{1,2,4,\dots,512\}$.
Using this configuration, the model has a receptive field of $N=2046$, which corresponds to about 46\,ms at the 44.1-kHz sample rate. The model response presented in Sec.~\ref{sec:bassman} suggests that a longer receptive field is needed for accurate predictions. However, experiments showed that increasing the receptive field did not significantly improve the performance with the validation data.

Two WaveNet models were selected for further evaluation. The small WaveNet (WaveNet1) uses 2 channels in the convolutional and post-processing layers whereas the other (WaveNet2) uses 16. These correspond to approximately 600 and 30,000 trainable parameters for the two models, respectively. WaveNets with more layers and channels were found to perform slightly better than the largest selected model, but due to the nature of the application, it was decided to keep the models to a reasonable size.

Different parameter combinations were also tested for the MLP, and the best one was selected for further evaluation. The chosen MLP has 8 hidden layers with 16 units each. The input layer size is set equal to the receptive field of the WaveNets. The total number of parameters is about 37,000.

\subsection{Objective evaluation}
\label{sec:obj-eval}
For objective evaluation, the ESR of the models was computed for a test set of unseen electric guitar and bass input signals. The unprocessed test set signals were obtained from Freesound, an online audio database \cite{font2013freesound}.
The ESR was computed for four conditions given by the combinations of low or high input level, and medium or high amplifier gain.
The input level refers to the voltage of the signal fed to the amplifier whereas the gain refers to the potentiometer setting in the preamplifier (medium: 50\%; high: 100\%).

\begin{table}[t!]
\caption{Error-to-signal ratio $\mathcal{E}$ for two gain settings and two input signal levels. The smallest errors are highlighted.}
\label{tab:obj}
\centering
\begin{tabular}{lllll}
\hline
Amp.~gain         & \multicolumn{2}{c}{50\%} & \multicolumn{2}{c}{100\%} \\ \hline
Input level     & Low        & High        & Low         & High        \\ \hline
Block \cite{eichas2016black}   & 3.4\%        & 2.0\%         & 15\%         & 12\%         \\
MLP      & 0.38\%        & 1.1\%         & 1.9\%         & 17\%         \\
WaveNet1 & 1.3\%        & 1.8\%         & 2.3\%         & 11\%         \\
WaveNet2 & \textbf{0.32\%}        & \textbf{0.49\%}         & \textbf{0.67\%}         & \textbf{2.0\%}
\end{tabular}
\end{table}

The results are shown in Table \ref{tab:obj}. In the 50\% gain setting conditions, the preamplifier's response is weakly nonlinear, and it can be seen that all models have a relatively small ESR. In the 100\% gain conditions, in which the preamplifier behaves more nonlinearly, ESR becomes larger. It can be seen that WaveNet2 significantly outperforms the other models in the high level and 100\% gain condition. This model also has the lowest ESR in all other conditions. Each model produced similar scores with guitar and bass signals such that only the combined results are shown in Table \ref{tab:obj}.

The offline processing speeds of the models are reported in Table~\ref{tab:speed}. Working with the 44.1-kHz sample rate, the Tensorflow implementations of the models run faster than real time (RT) on a 2.8-GHz Intel Core i5 CPU. This  suggests the models could be applicable for real-time processing.
\begin{table}[t]
\caption{Processing speed of the models on a CPU.}
\label{tab:speed}
\centering
\begin{tabular}{lllll}
\hline
Model               & Block & MLP & WaveNet1 & WaveNet2 \\ \hline
Speed ($\times$ RT) & 45    & 4.9   & 6.1        & 3.0       
\end{tabular}
\end{table}

When algorithm delay needs to be minimized for low-latency applications, the WaveNet can be run sample-by-sample, and in this case it becomes increasingly important to use fast inference algorithms. These algorithms exploit the dilated nature of the filters and use queues to avoid calculating same values multiple times \cite{Ramachandran2017-fast-generation-cnn, Arik2017-deepvoice}. The study of a buffered block-wise implementation of such a solution, applicable for low-latency processing, is left for future work.

\subsection{Listening test}
To assess the perceptual similarity between the models and the SPICE reference, a multiple stimuli with hidden reference and anchor (MUSHRA) listening test was conducted~\cite{schoeffler2018webmushra}. On each MUSHRA trial, the subjects were presented with a sound processed through the SPICE reference, and several test items which they were asked to rank based on how accurately they approximate the timbre of the reference.

The listening test items consisted of the models included in the objective evaluation. Additionally, a hidden reference, and an anchor obtained by processing the sounds through a hard clipper, were included on each trial. Five different input sounds were used: four bass guitar sounds and one electric guitar sound. The sounds were a subset of the samples used for the objective evaluation. The listening test focused on bass guitar sounds since they seemed to bring out the differences between the models most clearly. The reason may be that the bass sounds are monophonic and have a low fundamental frequency, so the harmonic distortion is well audible. Each of the five input sounds were processed with two settings: high input level and 50\% gain, and high input level and 100\% gain. This led to a total of ten trials for the MUSHRA test.

A total of 14 subjects participated in the test. The subjects were guitar and bass players with experience in critical listening. The test was conducted in sound-proof booths using Sennheiser HD-650 headphones. In post-screening, 4 subjects were excluded from the final analysis due to giving the hidden reference less than 90 points in two or more trials.

\begin{figure}[t]
  \centerline{\includegraphics{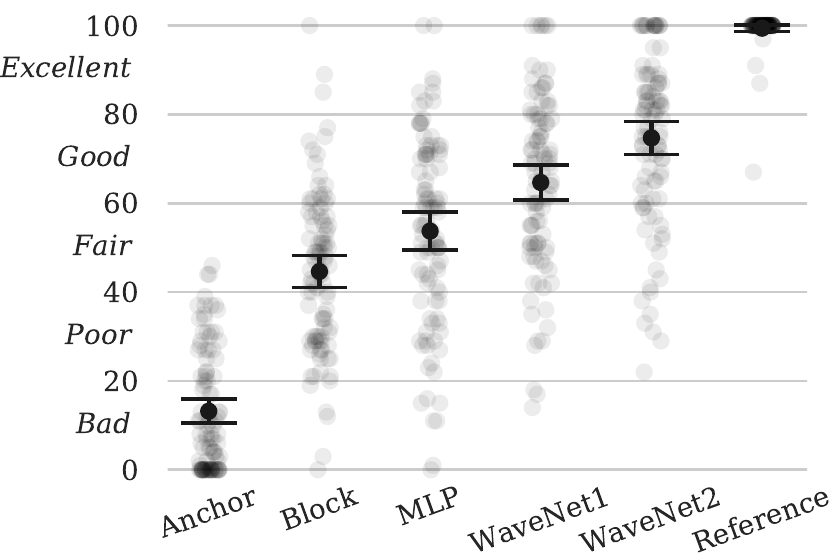}}
  \caption{Mean results of the MUSHRA listening test.}
  \label{fig:mushra}
\end{figure}

The results of the listening test are shown in Fig.~\ref{fig:mushra}. The black circles indicate the mean score for each method over all test items, and the error bars indicate the 95\% confidence interval. The transparent circles present the individual answers. There are significant differences between the subjective quality of the models. Overall, WaveNet2 performed best out of all the models. Related audio files are available online \cite{demopage}.

\section{Conclusions}
\label{sec:conclusions}
This work has discussed the use of deep learning for the emulation of nonlinear audio circuits with adjustable parameters. Pre-emphasis filtering has been necessary for successful training. Our tests suggest that the deep neural network can be run in real time at a typical audio sample rate. Our listening test results show that the proposed deep convolutional architecture outperforms a state-of-the-art black-box model. Future work will apply the proposed method to other devices, such as effect pedals with real training data.

\section{Acknowledgment} 
We acknowledge the computational resources provided by the Aalto Science-IT project.




\bibliographystyle{IEEEbib}
\bibliography{refs}

\begin{thebibliography}{10}

\bibitem{Karjalainen06}
M.~Karjalainen and J.~Pakarinen,
\newblock ``Wave digital simulation of a vacuum-tube amplifier,''
\newblock in {\em Proc. IEEE ICASSP}, Toulouse, France, May 2006, pp. 153--156.

\bibitem{Yeh2009}
D.~T. Yeh,
\newblock {\em Digital Implementation of Musical Distortion Circuits by
  Analysis and Simulation},
\newblock {PhD} thesis, Stanford University, Stanford, CA, June 2009.

\bibitem{Valimaki11}
V.~V{\"a}lim{\"a}ki, S.~Bilbao, J.~O. Smith, J.~S. Abel, J.~Pakarinen, and
  D.~Berners,
\newblock ``Virtual analog effects,''
\newblock in {\em DAFX: Digital Audio Effects}, U.~Z{\"o}lzer, Ed., pp.
  473--522. Wiley, Chichester, UK, second edition, 2011.

\bibitem{Barbour98}
E.~Barbour,
\newblock ``The cool sound of tubes,''
\newblock {\em IEEE Spectrum}, vol. 35, no. 8, pp. 24--35, Aug. 1998.

\bibitem{pakarinen2009review}
J.~Pakarinen and D.~T. Yeh,
\newblock ``A review of digital techniques for modeling vacuum-tube guitar
  amplifiers,''
\newblock {\em Computer Music J.}, vol. 33, no. 2, pp. 85--100, 2009.

\bibitem{dunkel2016fender}
W.~R. Dunkel et~al.,
\newblock ``The {F}ender {B}assman 5{F6-A} family of preamplifier
  circuits---{A} wave digital filter case study,''
\newblock in {\em Proc. Int. Conf. Digital Audio Effects (DAFx)}, Brno, Czech
  Republic, Sept. 2016, pp. 5--9.

\bibitem{kemp1999analysis}
M.~J. Kemp,
\newblock ``Analysis and simulation of non-linear audio processes using finite
  impulse responses derived at multiple impulse amplitudes,''
\newblock in {\em Proc. Audio Eng. Soc. 106th Conv.}, Munich, Germany, May
  1999.

\bibitem{farina2005emulation}
A.~Farina and E~Armelloni,
\newblock ``Emulation of not-linear, time-variant devices by the convolution
  technique,''
\newblock in {\em Proc. Congresso AES Italy}, Como, Italy, Nov. 2005.

\bibitem{helie2010volterra}
T.~H{\'e}lie,
\newblock ``Volterra series and state transformation for real-time simulations
  of audio circuits including saturations: Application to the {Moog} ladder
  filter,''
\newblock {\em IEEE Trans. Audio Speech Lang. Process.}, vol. 18, no. 4, pp.
  747--759, May 2010.

\bibitem{orcioni2018identification}
S.~Orcioni et~al.,
\newblock ``Identification of {Volterra} models of tube audio devices using
  multiple-variance method,''
\newblock {\em J. Audio Eng. Soc.}, vol. 66, no. 10, pp. 823--838, Oct. 2018.

\bibitem{tronchin2013emulation}
L.~Tronchin,
\newblock ``The emulation of nonlinear time-invariant audio systems with memory
  by means of {V}olterra series,''
\newblock {\em J. Audio Eng. Soc.}, vol. 60, no. 12, pp. 984--996, Dec. 2013.

\bibitem{wavenet2016oord}
A.~{van den Oord} et~al.,
\newblock ``{W}ave{N}et: A generative model for raw audio,''
\newblock {\em ArXiv pre-print}, 2016,
\newblock arXiv:1609.03499 {[cs.SD]}.

\bibitem{covert2013vacuum}
J.~Covert and D.~L. Livingston,
\newblock ``A vacuum-tube guitar amplifier model using a recurrent neural
  network,''
\newblock in {\em Proc. IEEE SoutheastCon}, Jacksonville, FL, Apr. 2013.

\bibitem{zhang2018vacuum}
Z.~Zhang et~al.,
\newblock ``A vacuum-tube guitar amplifier model using long/short-term memory
  networks,''
\newblock in {\em Proc. IEEE SoutheastCon}, Saint Petersburg, FL, Apr. 2018.

\bibitem{schmitz2018nonlinear}
T.~Schmitz and J.-J. Embrechts,
\newblock ``Nonlinear real-time emulation of a tube amplifier with a long short
  time memory neural-network,''
\newblock in {\em Proc. Audio Eng. Soc. 144th Conv.}, Milan, Italy, May 2018.

\bibitem{rethage2018wavenet}
D.~Rethage, J.~Pons, and X.~Serra,
\newblock ``A {W}avenet for speech denoising,''
\newblock in {\em Proc. IEEE ICASSP}, Calgary, Alberta, Canada, Apr. 2018, pp.
  5069--5073.

\bibitem{dempwolf2011physically}
K.~Dempwolf and U.~Z{\"o}lzer,
\newblock ``A physically-motivated triode model for circuit simulations,''
\newblock in {\em Proc. Int. Conf. Digital Audio Effects (DAFx)}, Paris,
  France, Sept. 2011, pp. 257--264.

\bibitem{fonseca2018general}
E.~Fonseca et~al.,
\newblock ``General-purpose tagging of {Freesound} audio with audioset labels:
  Task description, dataset, and baseline,''
\newblock {\em Submitted to DCASE2018 Workshop}, 2018,
\newblock arXiv:1807.09902 {[cs.SD]}.

\bibitem{diederik2014adam}
D.~P. Kingma and J.~Ba,
\newblock ``Adam: A method for stochastic optimization,''
\newblock in {\em Proc. Int. Conf. Learning Representations (ICLR)}, San Diego,
  CA, May 2015.

\bibitem{eichas2016black}
F.~Eichas and U.~Z{\"o}lzer,
\newblock ``Black-box modeling of distortion circuits with block-oriented
  models,''
\newblock in {\em Proc. Int. Conf. Digital Audio Effects (DAFx)}, Brno, Czech
  Republic, Sept. 2016, pp. 39--45.

\bibitem{eichas2017block}
F.~Eichas, S.~M{\"o}ller, and U.~Z{\"o}lzer,
\newblock ``Block-oriented gray box modeling of guitar amplifiers,''
\newblock in {\em Proc. Int. Conf. Digital Audio Effects (DAFx)}, Edinburgh,
  UK, Sept. 2017, pp. 5--9.

\bibitem{font2013freesound}
F.~Font, G.~Roma, and X.~Serra,
\newblock ``Freesound technical demo,''
\newblock in {\em Proc. 21st ACM Int. Conf. Multimedia}, Barcelona, Spain, Oct.
  2013, pp. 411--412.

\bibitem{Ramachandran2017-fast-generation-cnn}
P.~Ramachandran et~al.,
\newblock ``Fast generation for convolutional autoregressive models,''
\newblock in {\em Proc. Int. Conf. Learning Representations (ICLR)}, Toulon,
  France, Apr. 2017.

\bibitem{Arik2017-deepvoice}
S.~O. Arik et~al.,
\newblock ``{Deep Voice}: Real-time neural text-to-speech,''
\newblock in {\em Proc. Int. Conf. Machine Learning (ICML)}, Sydney, Australia,
  Aug. 2017.

\bibitem{schoeffler2018webmushra}
M.~Schoeffler et~al.,
\newblock ``web{MUSHRA}---{A} comprehensive framework for web-based listening
  tests,''
\newblock {\em J. Open Research Software}, vol. 6, no. 1, Feb. 2018.

\bibitem{demopage}
E-P. Damsk{\"a}gg, L.~Juvela, E.~Thuillier, and V.~V{\"a}lim{\"a}ki,
\newblock ``Deep learning for tube amplifier emulation,'' accompanying website,
\newblock available online at:
  http://research.spa.aalto.fi/publications/{\allowbreak}papers/{\allowbreak}icassp19-deep/.

\end{thebibliography}

\end{document}